\documentclass[runningheads]{llncs}

\usepackage[T1]{fontenc}
\usepackage{graphicx}

\usepackage{soul}

\usepackage{todonotes}

\usepackage{xurl}

\usepackage[utf8]{inputenc}

\begin{document}

\title{\st{Security-by-design} Securing a compromised system}

\author{Awais Rashid\inst{1}\orcidID{0000-0002-0109-1341} \and
Sana Belguith\inst{1}\orcidID{0000-0003-0069-8552} \and 
Matthew Bradbury\inst{2}\orcidID{0000-0003-4661-000X} \and
Sadie Creese\inst{3}\orcidID{0000-0002-2414-9657} \and 
Ivan Flechais\inst{3}\orcidID{0000-0002-3620-0843} \and  
Neeraj Suri\inst{2}\orcidID{0000-0003-1688-1167}}
\authorrunning{A. Rashid et al.}

\institute{University of Bristol, UK \and 
Lancaster University, UK \and 
University of Oxford, UK}


\maketitle

\begin{abstract}

Digital infrastructures are seeing convergence and connectivity at unprecedented scale. This is true for both current critical national infrastructures and emerging future systems that are highly cyber-physical in nature with complex intersections between humans and technologies, e.g., smart cities, intelligent transportation, high-value manufacturing and Industry 4.0. Diverse legacy and non-legacy software systems underpinned by heterogeneous hardware compose on-the-fly to deliver services to millions of users with varying requirements and unpredictable actions. This complexity is compounded by intricate and complicated supply-chains with many digital assets and services outsourced to third parties. The reality is that, at any particular point in time, there will be untrusted, partially-trusted or compromised elements across the infrastructure. Given this reality, and the societal scale of digital infrastructures, delivering secure and resilient operations is a major challenge. We argue that this requires us to move beyond the paradigm of security-by-design and embrace the challenge of \emph{securing-a-compromised-system}.

\end{abstract}

\section{Introduction}

The security of infrastructures, architectures, and mechanisms is built on assumptions. This includes speculations or approximations about context, usage, threat models or interactions with other systems. Even when correct, these assumptions may only hold at a particular point in time and are often shaped by additional assumptions about the system's lifespan and that the designed security approaches will mitigate against vulnerabilities over that lifespan. These assumptions do not survive contact with the reality of deployed systems. 

In practice, a system involves a range of other sub-systems several of which are not in the purview of the developers or the organisation deploying the system\textemdash in many instances assets and services are outsourced to third parties with security breaches having major knock-on effects on a wide array of systems and users~\cite{chowdhury2024a}. Even where these sub-systems are within the development or administrative control of the system owner, there are complex technology stacks with a plethora of third party libraries, hardware, software components and diverse development practices -- including often misplaced assumptions about threat models~\cite{chowdhury2024b,rashid2021}. Furthermore, threat actors evolve quickly in terms of their capabilities, motivations and tactics, techniques and procedures, for example using generative AI techniques to create malware~\cite{hp_report_2024}. Systems, particularly large-scale ones that underpin societal scale infrastructures, e.g., water, power, digital services for citizens, do not evolve as rapidly. It takes time and money to change a system and, where such change is enacted, for example, for a power system or railway infrastructures, it is a major investment of hundreds of millions or billions of pounds involving rearchitecting the system, upgrading hardware and software systems, testing for safety and uptime, and retraining of staff. In some cases, it is even not possible to upgrade legacy systems to state-of-the-art security mechanisms due to real-time requirements or the need to formally prove safety and dependability related properties. 

Furthermore, digital infrastructures have complex interdependencies and intersections with human users who are an integral part of the work and information flows. Often, human interactions with the systems catalyse dynamic composition of services which create new interactions and dependencies across systems at runtime. Usability of security mechanisms is paramount~\cite{cybok_hf} not only to ensure that security does not create significant overheads but also to mitigate against shadow security practices~\cite{kirlappos2015} by users. 
Security mechanisms typically aim to address specific threats or vulnerabilities. For example, the Digital Security by Design (DSbD) programme is making key advances to eliminate memory vulnerabilities at the hardware-level~\cite{dsbd}. While this holds great promise, there remain risks of developer-induced vulnerabilities~\cite{ullah2024} or constraining assumptions as to who the threat actor is, e.g., one aiming to extract data from RAM after super cooling it~\cite{wu2023}. 

The reality is that it is \emph{impossible} to secure all aspects of a system by design. Measuring security -- and its \emph{goodness} -- is an open problem and polling \emph{goodness} of a system cannot perfectly determine if the system's behaviour is good. The best one can do is probabilistic~\cite{bradbury2021}. The reality is that systems will become compromised or will always have untrusted, partially trusted, or compromised elements. Pragmatic considerations mean also that one cannot simply shutdown a whole transportation infrastructure because, say, a traffic signal is compromised, or disconnect large parts of the power grid because specific components are under attack. How we ensure that the system continues to operate within specified bounds of safety and resilience -- albeit potentially at reduced capacity -- is critical, as is the capability to limit impacts of partial breaches including cascading effects across interconnected infrastructures. \emph{We, therefore, posit that research needs to move beyond the paradigm of security-by-design and embrace the challenge of securing-a-compromised-system.} This requires scientific advances in four key dimensions. We discuss these next to present a research agenda for the research community. 

\section{Research Challenges}

\subsection{Predictability}

Predictability is an inherent goal in security: knowing what can and will happen, what can be done to mitigate it and the extent to which any mitigation is effective. Predictability requires \emph{measuring security} which is a hard problem in any system. It is compounded in digital infrastructures as complexity is paramount: mix of technology (legacy and non-legacy), uncertainty about threats and effectiveness of controls, emergent behaviour, interactions between security and other system goals, trustworthiness of people and organisations and divergence from rules (shadow practices).

A large body of work has focused on developing metrics. Reference sources such as NIST 800-55~\cite{swanson2003nist} and ISO 27004~\cite{international2009iso} adopt a catalogue approach: reference metrics classified into categories and documented with scenarios and examples. However, the contextualisation of metrics relies on arbitrary examples and use cases, limiting their expressiveness and hence their ability to address the complexity and inherent uncertainty. Others promote a more structured way of designing security measurements~\cite{herrmann2007complete,jaquith2007security,hayden2011itsec}. However, they presume that one knows a priori what is pertinent to measuring security and that instrumenting all elements is feasible\textemdash not the case given the dynamism and opaqueness in contemporary and future digital infrastructures. 

Standards such as NIST SP 800-160 Volumes 1 and 2~\cite{ross2021,ross2022} offer guidance on engineering trustworthy secure and resilient systems. However, such standards are based on the premise that the problem, solution and trustworthiness contexts can be established {\em a priori} and that systems can be architected with a high degree of control over their components. These assumptions do not hold in large-scale infrastructures. There are systems about which one can collect relevant metrics (e.g., a sub-system into which deep instrumentation can be deployed) and for others one can not. Uncertainty also comes from what is unseen, e.g., shadow practice. So modelling the dependencies and deriving relevant metrics to understand the security implications of those dependencies is a major scientific challenge.

\subsection{Composition}

Composing security provision in any system is a hard problem. For instance, a longstanding principle is that of \emph{secure distributed composition} which states that when multiple sub-systems or components are composed, the resulting system does not weaken the security policies enforced by its components. Security policy enforcement approaches typically take an organisation- or network-centric view of security, e.g.,~\cite{hadjiantonis2007,sloman1994}. These tend to be either obligation-driven or authorisation-driven~\cite{sloman1994}. In the former case, policies are enforced actions in response to particular events or stimuli within a system while, in the latter, they provide access control rules specifying whether a particular subject can legitimately access (or not) a particular object. Such approaches assume that the system, whether distributed or not, is within a single administrative control and even where platform or geographical boundaries are crossed, this happens within the control of a single organisation or a federated security management framework~\cite{decat2014}. This is not the case for digital infrastructures under discussion in this paper, which are globally interconnected open-ended networked environments.

The challenge is further compounded by the cyber-physical nature of many constituent systems where legacy hardware and software are abound and security assurances can vary widely\textemdash from poorly designed network protocol stacks to access control models that do not enforce privileges at suitable levels. Furthermore, such environments are not static. Devices, systems and services can dynamically (and, increasingly, automatically) compose based on context and locality. Human actors are integral to the dynamics, and often catalyse dynamic composition and delivery of services, e.g., through wearables that bridge multiple systems simultaneously. Consequently, security orchestration can be, at best, delivered through service-level agreements (SLAs). However, violation of such SLAs is often only detected post-hoc. Furthermore, in a large set of scenarios, e.g., those involving untrusted or partially-trusted third party systems, specification, agreement and enforcement of an SLA is impossible.

\subsection{Continual Assurance}

For well-structured systems (e.g., control systems, database/transactional systems) with clearly specified security requirements on a) interactions and dependencies across sub-systems, services and components, and b) the expected threats, research has developed a variety of sophisticated capabilities to monitor and analyse their security posture to assert (with varying levels of confidence and accuracy) the requisite levels of security assurances~\cite{ayodeji2023cyber,hudic2017security,macaulay2011cybersecurity}. This is not the case for globally interconnected open-ended networked heterogeneous environments where a complete awareness of all dependencies and knowledge of all operational paths is not viable. This becomes even more challenging in an ultra-large scale environment where conjunctions of secure and unsecure, trusted and untrusted, and reliable and
unreliable elements are present.

For instance, for complex and dynamically interconnected systems, the consequent lack of a) complete and stable system and security specifications including the threats, and b) complete and stable dependency and interface specifications, make provisioning of continual assurance a challenge. Such systems are typically heterogeneous couplings of structured, unstructured, synchronous and asynchronous elements and services. This precludes a single system model invariably considered in state-of-the-practice/art approaches~\cite{shukla2022system}.

\subsection{Incident Response}

Over the past 20 years significant progress has been made to mature and develop incident response and recovery capacity,  whether delivered by in-house security operations centres (SOCs) or by third party managed service providers.  This is supported by automation and tooling, often in the form of Security Information and Event Management (SIEM) systems that provide real-time information to human operators in a SOC. However, selecting the best response and recovery actions remains a largely human task~\cite{bada2014computer}. Orchestrating incident response on an infrastructure-scale requires research into the appropriate balance between human-machine decision-making.

Existing standards such as ISO/IEC 27035-2:2023~\cite{iso27035} offer guidelines on how to plan, prepare and learn lessons from any incidents, both in terms of system defences and the incident response approach. Given the high-level nature of such guidance, operationalisation happens through \emph{playbooks}, acting as recipes on steps and actions to take during incident response. However, playbooks remain very much a manual setup, often taking the format of natural language texts or flow charts\textemdash typically in printed format placed in SOCs. Recent works have argued for more systematic model-based representations of playbooks~\cite{shaked2022}, and have highlighted the lack of a) usability studies of playbooks, and b) specificity even for highly rated playbooks for completeness and correctness by experts~\cite{stevens2022}.

In the infrastructures under discussion, each constituent system will have its own playbook unlikely to be formalised into any structured or systematic common model~\cite{shaked2022}. Orchestrating a globally coordinated incident response on this scale is, therefore, a major research challenge. It is made even more challenging by the dynamism\textemdash systems composing with the infrastructure or leaving. Furthermore, constituent systems' playbooks will change in response to incidents over time. So one cannot start from the assumption that the playbooks are convergent or will remain so over time. The complexity is further compounded because contextual information is a challenge in SIEMs as SOC workers are not involved in the design choices, configurations and operation of specific organisational assets from where telemetry is fed into the SOC\@. Where contextual information is communicated, this happens informally and thus remains tacit and not formally documented~\cite{bhatt2014}. 

\section{In Conclusion}

Advancing the paradigm of \emph{securing-a-compromised-system} will require a \emph{systems} approach that addresses the aforementioned four dimensions. We need new ways to elicit, specify, and validate security assurances for service composition in the presence of uncertainty, dynamism, and human behaviour. New mechanisms to compose and orchestrate security provision across diverse and heterogeneous evolving infrastructures with legacy and non-legacy elements will be critical in this regard. Alongside, it is paramount that the research community develops ways to reason about the security state at runtime in order to provide continuity of oversight and trust in the presence of partially trusted, under attack, vulnerable, or compromised elements. Last, but by no means least, it is essential that we address how we may orchestrate incident response that accounts for heterogeneous incident response practices in constituent systems and provides situational awareness at the necessary pace and resolution for optimal human-machine decision-making. 

\begin{credits}
\subsubsection{\ackname} This research is supported by the Engineering and Physical Sciences Research Council grant SCULI: Securing Convergent Ultra-large Scale Infrastructures [EP/Z531315/1].

\end{credits}

\bibliographystyle{splncs04}
\bibliography{bibliography}

\end{document}